# Monopoles and Quark Confinement: Introduction and Overview[*]

Ken Yee

*Department of Physics and Astronomy, L.S.U.*
*Baton Rouge, Louisiana   70803-4001, USA*

e-mail: kyee@rouge.phys.lsu.edu

February 20, 1994

## Abstract

We (try to) pedagogically explain how monopoles arise in QCD, why maximal Abelian(MA) gauge is "special" for monopole study, the Abelian projection in MA gauge, its resultant degrees of freedom(photons, monopoles and charged matter fields), species permutation symmetry, and the QCD-equivalent action in terms of these degrees of freedom. Then we turn to more recent developments in the subject: Abelian dominance, large $N$ behavior of Abelian projected QCD, mass of the charged matter fields, notion of an effective photon-monopole action obtained by integrating out the charged matter fields, and problems encountered in evaluating this effective action using the microcanonical demon method on the lattice.

---

## 1. Abelian Projection of QCD

An open problem in QCD is to identify the quark confinement mechanism and understand how it works. To this end compact or lattice QED(CQED), whose action is $-S_{CQED} = \sum_{\mu<\nu} \beta_{CQED} \cos\Theta_{\mu\nu}$ provides a compelling prototype. In lattice different forms notation [1], the expectation value of a Wilson loop $W \equiv \exp i(A, J)$ in CQED upon a BKT transformation [2] is

$$\langle W \rangle \propto \sum_{\{k|\partial k=0\}} \exp\{-\widetilde{S}\}, \qquad (1)$$

where

$$\widetilde{S} \equiv \frac{1}{2\beta_{CQED}}(J, \Delta^{-1}J) + 2\pi^2 \beta_{CQED}(k, \Delta^{-1}k) - 2\pi i(^*dk, \Delta^{-1}E). \qquad (2)$$

The 1-forms $J$ and $k$ are, respectively, conserved electric and magnetic monopole current loops. 2-form $E$ is the electromagnetic field due to external current $J$: $\partial E = J$. $\Delta^{-1}$ is the inverse Laplacian. The first and second terms of $\widetilde{S}$ correspond to the electromagnetic interaction energies of $J$ and $k$. The third term is the interaction between the monopole currents and the background electric field $E$ created by $J$.

In the Meissner effect of BCS superconductivity, copper pairs—bosons carrying *electric* charge—dynamically squeeze magnetic flux into tubes which act to confine magnetic charges. When coupling $\beta_{CQED}$ is sufficiently small, the entropy of the sum over monopole loops in (1) dominate over suppression by Boltzmann factor $\exp\{-\widetilde{S}\}$ and monopoles are said to be "condensed." In this phase CQED exhibits the *dual* Meissner effect. Simulations indicate that the CQED vacuum looks like an effective dual Type II superconductor [3]: *magnetic* monopoles, responding to the background electric field $E$, rearrange the electric field so that there is a net electric flux tube between the Wilson loop. The energy per unit length of this flux tube is the string tension. In this way, magnetic monopole condensation causes electric confinement in CQED.

This characterization that monopoles are condensed in the confinement phase is formally justified as follows. CQED can be mapped to an Abelian Higgs model [4]. The shape of the effective potential $V(\phi)$ governing the Higgs field $\phi$, which is closely related to the monopole creation operator, depends on the phase of CQED. In the confining phase $V(\phi)$ has a minimum at $\phi \neq 0$ and, accordingly, vacuum expectation value $\langle\phi\rangle \neq 0$. Thusly, monopoles are condensed in CQED's confinement phase. In the deconfined phase, $\langle\phi\rangle = 0$.

An analogous demonstration that monopole condensation is the origin of QCD confinement would be a great achievement [5]. But where are the monopoles in QCD? 't Hooft suggested the following idea [6]. Suppose QCD monopoles, like the 't Hooft-Polyakov monopoles of the Georgi-Glashow model [7], carry charges that are magnetic with respect to the $[U(1)]^{N-1}$ Cartan subgroup of color $SU(N)$. Then $SU(N)$ gauge symmetry obscures the magnetic charges and it is necessary to gauge fix at least the $SU(N)/[U(1)]^{N-1}$ symmetry to expose them.

To this end, let $\mathbf{X}$ be a hermitian, traceless adjoint field transforming locally as

$$\mathbf{X}(x) \to \Omega(x)\mathbf{X}(x)\Omega^\dagger(x). \tag{3}$$

Consider the gauge in which **X** is diagonalized and its eigenvalues ordered according to increasing size. Such a gauge is achievable on any background gauge field because **X** transforms locally under $\Omega$. Except at sites where **X** has degenerate eigenvalues, this condition fixes the gauge completely modulo diagonal $[U(1)]^{N-1}$ gauge transformations

$$\Omega_{\text{residual}} = \begin{pmatrix} \exp^{-i\omega_1} & & \\ & \ddots & \\ & & \exp^{-i\omega_N} \end{pmatrix}, \quad \sum_{i=1}^{N} \omega_i = 0. \tag{4}$$

Would-be QCD monopoles might arise as follows. Suppose **X** has two degenerate eigenvalues at $x_o$ in a $2 \times 2$ block **x** of **X**. In the neighborhood around $x_o$, **x** would be a $2 \times 2$ hermitian matrix

$$\mathbf{x}(x) = \Phi_o(x)\mathbf{1} + \sum_{i=1}^{3} \sigma_i \Phi_i(x). \tag{5}$$

The $\Phi_i$ are real functions and $\sigma_i$ the Pauli spin matrices. Eigenvalue degeneracy at $x_o$ means $\Phi_1(x_o) = \Phi_2(x_o) = \Phi_3(x_o) = 0$. In $D = 3+1$ dimensional spacetime the typical loci of points simultaneously obeying these three conditions are *lines*. Assuming **X** is an analytic field, Taylor expansion yields

$$\Phi_i(x) = (x - x_o) \cdot \nabla \Phi_i(x_o) + \mathcal{O}(x - x_o)^2 \qquad i = 1, 2, 3. \tag{6}$$

$\Phi_i$ near $x_o$ is (up to coordinate stretching) a "hedgehog" field and, in spherical coordinates centered at $x_o$, the $SU(2)$ gauge transformation which diagonalizes a hedgehog field is [7]

$$\Omega(x) = \begin{pmatrix} \cos\frac{\Theta}{2} & \exp^{-i\phi} \sin\frac{\Theta}{2} \\ -\exp^{i\phi} \sin\frac{\Theta}{2} & \cos\frac{\Theta}{2} \end{pmatrix}. \tag{7}$$

$\Omega(x_o)$ is ill-defined but it does not violate the gauge condition, which is ambiguous at $x_o$ since $\mathbf{x}(x_o) \propto \mathbf{1}$. Under gauge transformation

$$A_\mu \to \Omega(A_\mu + \frac{i}{g}\partial_\mu)\Omega^\dagger \tag{8}$$

the $SU(2)$ gauge field inside the $2 \times 2$ subspace gains a component

$$A_\phi^3 = \frac{i}{gr\sin\theta}\left(\Omega\partial_\phi\Omega^\dagger\right)^3 = \frac{1-\cos\theta}{2gr\sin\theta}. \tag{9}$$

This is the field of a monopole carrying magnetic charge proportional to $(+1, -1)$ with respect to the $U(1)$ subgroup generated by $\sigma_3$ within the $2 \times 2$ subspace. Hence, the lines where $\mathbf{X}$ has degenerate eigenvalues correspond to worldlines of monopoles carrying charge proportional to

$$(\cdots, 0, +1, -1, 0, \cdots). \tag{10}$$

These charges are magnetic with respect to the $[U(1)]^{N-1}$ residual gauge symmetry.

Whether these monopoles are condensed or not in the QCD vacuum depends on both the choice of gauge fixing operator $\mathbf{X}$ and the nature of the gauge configurations dominating the QCD path integral. 't Hooft conjectured that, in fact, for a right choice of $\mathbf{X}$ these monopoles are manifestations of gauge field features responsible for QCD confinement. These features *appear* as magnetic monopoles in certain gauges. In these gauges one can hope to have a fixed-gauge picture of QCD confinement caused by monopole condensation. In other gauges the underlying gauge field features causing confinement are still present, but they do not appear as monopoles.

The nonperturbative nature of this conjecture requires calculations that were thought to be prohibitively hard until it was realized that relevant numerical calculations are feasible in lattice QCD [8]. Yet, as there is no elementary or otherwise natural candidate for **X**, it was not clear which gauge to use. It turns out [9]-[17] a compelling gauge is maximal Abelian(MA) gauge. Upon decomposing gauge field $A$ into purely diagonal($n$) and purely off-diagonal($ch$) parts

$$A = A^n + A^{ch}, \tag{11}$$

MA gauge is

$$D_\mu^n A_\mu^{ch} \equiv \partial_\mu A_\mu^{ch} - ig[A_\mu^n, A_\mu^{ch}] = 0. \tag{12}$$

While MA gauge is a differential rather than an **X**-diagonalization condition, it similarly leaves a residual $[U(1)]^{N-1}$ symmetry, Eq. (4). Under $\Omega_{\text{residual}}$ the $N$ diagonal matrix elements $(A^n)_{ii}$ transform as neutral photon fields whereas the $N(N-1)$ offdiagonal matrix elements $(A^{ch})_{ij}$ transform as charged matter fields:

$$(A_\mu^n)_{ii} \to (A_\mu^n)_{ii} - \frac{1}{g}\partial_\mu \omega_i, \tag{13}$$

$$(A_\mu^{ch})_{ij} \to (A_\mu^{ch})_{ij} \exp^{-i(\omega_i - \omega_j)} \qquad i \neq j, \quad i,j \in [1,N]. \tag{14}$$

Since $(A^{ch})_{ij}$ carries two different $U(1)$ charges, the $A^{ch}$ fields induce "interspecies" interactions between the $N$ photons.

MA gauge can be motivated [17] by considering the $SU(N)$ Georgi-Glashow(GG) model, which has an adjoint, bare mass $M$ Higgs field $\Phi$ coupled gauge invariantly to $A$. We can think of (pure) QCD as being the formal $M \to \infty$ limit of the GG model because $\Phi$ freezes out and decouples in this limit. At all $M$, GG has finite energy 't Hooft-Polyakov monopole solutions magnetic according to electromagnetic field tensor [7]

$$f_{\mu\nu} = \partial_\mu(\hat{\Phi}^a A_\nu^a) - \partial_\nu(\hat{\Phi}^a A_\mu^a) + \frac{i}{g}\hat{\Phi}^a[\partial_\mu\hat{\Phi}, \partial_\nu\hat{\Phi}]^a \qquad (15)$$

where $\hat{\Phi}^a = \Phi^a/(\Phi^b\Phi^b)^{\frac{1}{2}}$. The value of $f_{\mu\nu}$ is gauge invariant but the three terms on the RHS of (15) mix under gauge transformations. The evaluation of $f_{\mu\nu}$ simplifies in gauges in which one or two of the three terms on the RHS of (15) vanish. MA gauge can be defined as the gauge in which $\Phi$ is diagonalized. Diagonalization of $\Phi$ induces a gauge transformation on the *monopole solutions* so that they obey MA gauge condition (12). In this gauge $f_{\mu\nu}$ for monopole fields reduces to the Abelian form

$$f_{\mu\nu} \equiv \partial_\mu A_\nu^n - \partial_\nu A_\mu^n. \qquad (16)$$

As $\Phi$ is undefined in (pure) QCD it is unclear how to use (15) to identify magnetic monopoles in QCD. However, (12) and (16) do not depend explicitly on $\Phi$. This fortuitous fact allows one to try and identify monopoles in QCD by fixing the gauge fields to (12) and, following (16), evaluating $f_{\mu\nu}$ by treating the $(A^n)_{ii}$ as Abelian fields. On the lattice the monopole currents are identified according to a discretized version of[1]

$$k_\mu \equiv \frac{1}{2\pi}\epsilon_{\mu\nu\lambda\delta}\partial_\nu f_{\lambda\delta}, \qquad (17)$$

a procedure known to be appropriate for CQED [18]. On the lattice $A_\mu^n$ is compact—$(A_\mu^n)_{ii} \in [-\pi, \pi)$—so that, as in CQED, it potentially may have nonzero magnetic monopole currents. Note that since $\partial_\mu k_\mu = 0$ by definition of $k_\mu$, monopole currents always flow in closed loops.

---
[1] Our definition is a factor of 2 different from another common normalization, $k_\mu \equiv \frac{1}{4\pi}\epsilon_{\mu\nu\lambda\delta}\partial_\nu f_{\lambda\delta}$.

This procedure where only the diagonal $A^n$ components of nonAbelian gauge fields $A$ are used to determine the monopole-related electromagnetic fields is called *Abelian projection*(AP). As operational exercises people have performed AP starting from a variety of gauges. As anticipated, the results vary with gauge. Only MA gauge has emerged as promising. While this certainly does not preclude the existence of some as-yet untried better gauge, all other tested trial gauges lead to at least one bad consequence which rules it out.

For $SU(2)$ QCD the following results hold in MA gauge: monopoles have a nonzero number density which persists as the lattice spacing is taken smaller and smaller [10]; they are quantifiably more dynamical in the confining phase than the finite temperature deconfined phase [8, 11]; their density seems to correlate to the nonAbelian string tension under cooling [13]; reminiscent of cooper pairs in the Meissner effect, the monopole currents circulate around effective chromoelectric flux tubes [14]; in the finite temperature deconfined phase the monopole density does *not* vanish, as they would not if they are also responsible for the string tension of spatial Wilson loops [15]. Some of these $SU(2)$ results have been independently verified by the author for $SU(3)$ [16]. 't Hooft's conjecture seems to be supported.

In the remainder of this Section we show that interspecies interactions are $1/N$ suppressed. This indicates that the matter fields $A^{ch}$, which mediate interspecies interactions by virtue of their two-species charges, lose their influence at large $N$. Since $\sum_{i=1}^{N}(A_\mu^n)_{ii}$ is invariant under (13), an irreducible representation of $[U(1)]^{N-1}$ is

$$\theta_\mu^i \equiv (A_\mu^n)_{ii} - \Lambda_\mu, \qquad \Lambda_\mu \equiv \frac{1}{N}\sum_{j=1}^{N}(A_\mu^n)_{jj}. \tag{18}$$

While vector field $\Lambda$ is $[U(1)]^{N-1}$ invariant, the $\theta^i$ transform as $\theta_\mu^i \to \theta_\mu^i - \frac{1}{g}\partial_\mu \omega_i$ and obey constraint

$$\sum_{i=1}^{N} \theta_\mu^i = 0. \tag{19}$$

We shall refer to the quantum dynamics of the $N$ angles $\theta^i$, which comprise a compact $[U(1)]^{N-1}$-invariant gauge field theory, as Abelian projected QCD or APQCD. As described in Section 2, APQCD is the field theory obtained by integrating out $A^{ch}$ and $\Lambda$ from QCD in MA gauge. The dynamical variables of such Abelian gauge theories generically are photons, magnetic monopole current loops, and virtual electric current loops [19]. Due to (19), the AP electromagnetic field tensors $f_{\mu\nu}^i \equiv \partial_\mu\theta_\nu^i - \partial_\nu\theta_\mu^i$ obey $\sum_{i=1}^{N} f_{\mu\nu}^i = 0$ and, because monopoles always occur in charge-anticharge partners a la Eq. (10),

$$\sum_{i=1}^{N} k_\mu^i = 0. \tag{20}$$

APQCD expectation values have a species permutation symmetry by which [20] every species is equivalent to every other species; for $i \neq j$ and $i \neq l$ the relationship of species $i$ to $j$ is the same as $i$ to $l$. If $\mathcal{A}^i$ and $\mathcal{B}^j$ refer to two operators $\mathcal{A}$ and $\mathcal{B}$ composed exclusively of species $i$ and $j$ links, species permutation implies that

$$\langle \mathcal{A}^i \mathcal{B}^i \rangle = \langle \mathcal{A}^j \mathcal{B}^j \rangle, \qquad \langle \mathcal{A}^i \mathcal{B}^j \rangle = \langle \mathcal{A}^i \mathcal{B}^k \rangle, \quad j \neq i,\; k \neq i. \tag{21}$$

There is no implicit summation over repeated species indices in Eq. (21).

Let $c^i$ be any operator such as $\theta^i$, $f^i_{\mu\nu}$, or $k^i_\mu$ which obeys

$$\sum_{i=1}^{N} c^i = 0. \tag{22}$$

Together with species permutation symmetry (22) implies that

$$\langle c^i \rangle = -\sum_{j \neq i} \langle c^j \rangle = -(N-1)\langle c^j \rangle, \tag{23}$$

which in turn implies that $\langle c^i \rangle = 0$. (21) and (22) also imply

$$\langle \mathcal{A}^j \, c^k \rangle = -\left(\frac{1}{N-1}\right)\langle \mathcal{A}^i \, c^i \rangle \quad j \neq k. \tag{24}$$

(24) says the correlator between two different species is $1/N$ suppressed relative to the same correlator between the same two operators of the same species. Interspecies interactions are $1/N$ suppressed and in the large $N$ limit the $N$ species decouple.

What does (24) tell us about confinement? Consider

$$\overline{c}(i,j) \equiv -i\,\frac{\langle W^j \, c^i \rangle}{\langle W^j \rangle} \tag{25}$$

where $W^j$ is the $j^{\text{th}}$-species time-like abelian Wilson loop (see Eq. (28) below) which we take to be suitably much larger than the abelian flux tube width. $\overline{c}(i,j)$ is the expectation value of operator $c^i$ in the background electric field created by a widely-separated static $(q\overline{q})^j$ pair. Eq. (24) implies that

$$\overline{c}(i,j) = -\left(\frac{1}{N-1}\right)\overline{c}(j,j) \quad i \neq j. \tag{26}$$

A physical interpretation emerges if, for example, we set $c^i = E^i$, the $i^{\text{th}}$-species electric field. (26) implies the effective electric field $\overline{E}(i,j)$ points in the opposite direction of $\overline{E}(j,j)$ and that $\overline{E}(i,j)$ is suppressed relative to $\overline{E}(j,j)$ by $\frac{1}{N-1}$. The effective Abelian electric fields created by a $(q\overline{q})^j$ pair have a tendency to anti-align!

## 2. APQCD Action and Integrating Out $A^{ch}$

How do the Abelian $k^i$ monopole currents cause string tension in the *nonAbelian* Wilson loops? This is prima facially a difficult question, and it is not at all obvious (or even likely) that a CQED-like picture is applicable even if the Abelian projection correctly identifies the monopoles. In MA gauge the QCD Lagrangian $\mathcal{L}^{QCD} = -\frac{1}{2}\sum_{\mu\nu} \text{tr} F_{\mu\nu}^2$ is decomposable as

$$\mathcal{L}^{QCD} = -\frac{1}{2}\sum_{\mu\nu} \text{tr}\left(f_{\mu\nu}^2 + V_{\mu\nu}^2 - g^2 T_{\mu\nu}^2 - 2ig(f_{\mu\nu} + V_{\mu\nu})T_{\mu\nu}\right) \qquad (27)$$

where $V_{\mu\nu} \equiv D_\mu^n A_\nu^{ch} - D_\nu^n A_\mu^{ch}$, $T_{\mu\nu} \equiv [A_\mu^{ch}, A_\nu^{ch}]$, and $f_{\mu\nu}$ is defined in (16). The second and fourth terms in the RHS of (27) contain interactions between the neutral $A^n$ (or equivalently the $\theta$ and $\Lambda$) fields and the charged $A^{ch}$ fields. Further, according to the second, third, and fourth terms the $A^{ch}$ fields propagate and self-interact. Hence, not only does the nonAbelian Wilson loop $W \equiv P \exp i(A, J)$ contain $A^n$ and $A^{ch}$ components mixed together in a complicated way, the magnetic fields of the $A^n$ monopoles must penetrate through a QCD vacuum populated with virtual $A^{ch}$ loops.

To fix ideas, consider a simpleminded scenario in which the nonAbelian Wilson loop is dominated by its Abelian components, that is,[2]

$$\text{tr}\langle W \rangle \longmapsto \lambda \sum_{i=1}^{N} \langle W^i \rangle, \qquad W^i \equiv \exp i(J^i, \theta^i). \qquad (28)$$

---

[2] In this Section we always assume QCD has been fixed to MA gauge. Since (28) relies on decomposition (11), it is unambiguous only if the $SU(N)/[U(1)]^{N-1}$ gauge symmetry is fixed. Abelian Wilson loops $W^i$ are invariant under only $[U(1)]^{N-1}$ and not the full $SU(N)$.

where $\lambda$ is some proportionality parameter and "$\longmapsto$" means equality only in the very large Wilson loop limit. According to (28), the nonAbelian string tension is given by the string tensions of the $N$ Abelian Wilson loops $\langle W^i \rangle$, which are all the same by species permutation symmetry. I do not know a formal justification for (28). Numerically, in $SU(2)$ simulations Abelian Wilson loops seem to reproduce the nonAbelian string tension [11, 12], a result called "Abelian dominance" by its discoverers.

Assuming (28) has some truth in it, let us consider where it leads. According to Eq. (24), current $k^i$ correlates to loop $W^j$ so that the $W^j$ string tension is affected not only by $k^j$ monopoles but also $k^i$ ($i \neq j$) monopoles. Thus, even assuming (28) the situation is more complex than CQED: the $W^j$ string tension has contributions from not only $k^j$ but also $k^i$. Photons and $A^{ch}$ mediate the cross-species interactions.

If we are interesting in just long distance confinement physics, we might seek a simplification by anticipating that the $A^{ch}$ fields have nonzero mass $M_{ch}$. At distance scales longer than $1/M_{ch}$, we can integrate out the $A^{ch}$ fields and formulate QCD confinement exclusively in terms of the $A^n$ fields, which hypothetically contain the confinement-causing monopoles in the first place. Then we might hope to understand Abelian string tension as due to the action of monopoles and photons without the complication of virtual $A^{ch}$ loops.

$M_{ch}$ is estimated as follows. As is well-known [25], the nonAbelian adjoint Wilson loop crosses over from an area to a perimeter law beyond some critical size because a virtual $AA^{\dagger}$ pair pops out of the vacuum once the en-ergy stored in the $q\bar{q}$ string exceeds the pair mass, which is roughly twice the effective gluon mass [26]. The Abelian projection image of this phenomenon occurs when the $\langle W^i W^{j\dagger} \rangle$ string pops an $A^{ch} A^{ch\dagger}$ pair out of the vacuum. In $SU(3)$ the effective gluon mass is of order $M_g \sim 600 MeV$. This value, obtained from the pole of the gluon propagator [28], is *not* a selfevident definition of gluon mass. Indeed, $M_g$ varies with gauge [29, 30]. (It is not inconsistent for the pole of the gluon propagator to vary with gauge since, because of confinement, gluon mass is not a direct observable.) If $\langle W^i W^{j\dagger} \rangle$ crosses over to perimeter law at the same Wilson loop size as the nonAbelian adjoint Wilson loop and Abelian dominance extends to adjoint Wilson loops, then the $A^{ch}$ mass also must be of order

$$M_{ch} \sim M_g \sim 600 MeV. \qquad (29)$$

We stress that (29) is only a heuristic estimate; a numerical study of $M_{ch}$ is currently in progress [27].

Formally integrating out the charged matter fields yields [21]

$$-S_{APQCD}[\theta^1, \cdots, \theta^N] \equiv \log\Big\{\int [dA^{ch} d\Lambda] \ \exp(-S_{QCD}) \ \delta[D^n_\mu A^{ch}_\mu]\Big\}, \qquad (30)$$

where we have reexpressed $A^n$ in terms of the $\theta^i$. We have also integrated out $\Lambda$ which, being a $[U(1)]^{N-1}$ singlet, is not a gauge field. $S_{APQCD}$ is a $[U(1)]^{N-1}$ invariant action in which monopoles arise as topological quantum fluctuations in the compact fields $\theta^i$. Of course, there is no guarantee that $S_{APQCD}$ has a simple form or is otherwise well-behaved. However, if it is and one is able to obtain an expression for $S_{APQCD}$, one can apply the CQED techniques [2, 4] to analyse APQCD. This potentially would lead to an unambiguous demonstration that QCD monopoles are condensed, and a dynamical picture of how they cause Abelian string tension.

There are several possible representations for an action with $[U(1)]^{N-1}$ gauge invariance and monopoles [21, 22, 23, 24]. Since we evaluate $S_{APQCD}$ couplings on the lattice, the most suitable for us is an extension of lattice QED to $N$ interacting $U(1)$ species. We will focus on $N = 3$; extension to larger $N$ is straightforward. One operator obeying gauge invariance and species permutation symmetry is[3]

$$\sum_{i=1}^{3} \sum_{q=1}^{\infty} \beta_q \cos q f_{\mu\nu}^i. \qquad (31)$$

My numerical calculations(described below) indicate that $\beta_1 >> \beta_{q>1} \sim 0$ and, in general, $q = 1$ operators have substantially bigger $S_{APQCD}$ couplings than their $q > 1$ counterparts. This is plausibly because $S_{QCD}^{\text{lattice}}$ itself contains only plaquettes in the fundamental representation and the Abelian angles $\theta^i$ are faithfully imbedded in the gauge fields $A$. In addition to $1 \times 1$ plaquette $\cos q f_{\mu\nu}$ one might also consider $L \times M$ Wilson loops. Numerical simulations (see below) indicate that these larger Abelian Wilson loops are essentially absent from $S_{APQCD}$, possibly because $S_{QCD}^{\text{lattice}}$ is comprised of only $1 \times 1$ plaquettes.

Therefore, let us momentarily consider only $q = 1$, $1 \times 1$ loops. In addition to some functional of monopole currents $k^i$ and expression (31), the only two other possible quasi-local, gauge invariant operators are

$$\cos(f_{\mu\nu}^i + f_{\mu\nu}^j), \quad \cos(f_{\mu\nu}^i - f_{\mu\nu}^j) \qquad i \neq j. \qquad (32)$$

---

[3]$q$ must be an integer for $\cos q\theta^i$ to be $U(1)$ gauge invariant.

Since $f_{\mu\nu}^1 + f_{\mu\nu}^2 = -f_{\mu\nu}^3$ by (19), $\cos(f_{\mu\nu}^i + f_{\mu\nu}^j)$ is already included in (31). On the other hand, $\cos(f_{\mu\nu}^i - f_{\mu\nu}^j)$ is not included, but numerical simulations indicate that their couplings vanish in $S_{APQCD}$. Perhaps this is because $\cos(f_{\mu\nu}^1 - f_{\mu\nu}^2) = \cos(2f_{\mu\nu}^1 + f_{\mu\nu}^3)$ which contains a $q = 2$ component. Hence, a close approximation to $S_{APQCD}$ is

$$-S_0 = \log \delta_\theta + \log \delta_k + \sum_{i=1}^{3}\left\{-\frac{\kappa}{2}(k^i, k^i) + \beta \sum_{x,\mu<\nu} \cos f_{\mu\nu}^i\right\} \quad (33)$$

$$= \log \delta_k - \kappa[(k^1, k^2) + \sum_{i=1}^{2}(k^i, k^i)] + \beta \sum_{x,\mu<\nu} [\cos(f_{\mu\nu}^1 + f_{\mu\nu}^2) + \sum_{i=1}^{2} \cos f_{\mu\nu}^i]$$

where $\delta_\theta$ and $\delta_k$ are delta functions which enforce (19) and (20). (On the lattice (19) does not automatically imply (20) so each requires its own delta function.) In (33) we have allowed for a $\beta$-independent monopole mass parameter $\kappa$ a la Ref. [24]. If $S_0$ accurately models $S_{APQCD}$, we can prove that monopole condensation causes confinement in APQCD: a BKT transformation [2] of the $S_0$ partition function yields

$$\int [d\theta_\mu^i] \exp^{-S_0} \longmapsto \sum_{\{k_\mu^1, k_\mu^2 | \partial_\mu k_\mu^i = 0\}} \exp^{-S_{mono}} \quad (34)$$

where

$$S_{mono} = \left(k^1, (\kappa + 4\pi^2\beta\Delta^{-1})k^2\right) + \sum_{i=1}^{2}\left(k^i, (\kappa + 4\pi^2\beta\Delta^{-1})k^i\right). \quad (35)$$

The phases of (34) are determined by monopole condensation.

To examined how well $S_0$ corresponds to APQCD, first we generate an ensemble of importance sampling APQCD gauge configurations by applying the Abelian projection to an ensemble of Monte Carlo lattice QCD configurations. We seek the $[U(1)]^2$ action, $S_{APQCD}$, which would generate the same ensemble of APQCD configurations. To this end, we introduce an ansatz for $S_{APQCD}$ and use the microcanonical demon technique [31] to determine the optimal coupling constants of that ansatz. If the ansatz contains all the operators of $S_{APQCD}$ the microcanonical demon technique measures all the coupling constants *exactly* up to statistical errors. In practice, however, we apply the technique only to simple truncated actions which are unlikely to contain all $S_{APQCD}$ operators. If an operator is missing, the microcanonical demon gives effective values for the ansatz couplings adjusted to optimally fit the ensemble. These effective values would not be the same as the true values if all operators are included.

Table 1: APQCD couplings $\beta_q(L)$ for trial action $S_a$.

| $\beta_{QCD}$ | $\beta_1(1)$ | $\beta_2(1)$ | $\beta_3(1)$ | $\beta_1(2)$ | $\beta_2(2)$ | $\beta_3(2)$ |
|---|---|---|---|---|---|---|
| 5.7 | .82(.04) | .083(.005) | -.004(.004) | -.026(.002) | .001(.006) | -.01(.01) |
| 6.0 | .87(.02) | .133(.002) | -.013(.007) | -.052(.002) | .008(.004) | -.009(.002) |

Table 2: APQCD couplings $\beta_q(L=1)$ and $\kappa$ for trial action $S_b$.

| $\beta_{QCD}$ | $\beta_1(1)$ | $\beta_2(1)$ | $\beta_3(1)$ | $\kappa$ |
|---|---|---|---|---|
| 5.7 | .82(.02) | .077(.001) | -.007(.004) | -.028(.006) |
| 6.0 | .77(.02) | .129(.002) | -.021(.001) | -.058(.002) |

Table 1 lists the results for ansatz

$$-S_a \equiv \log \delta_\theta + \log \delta_k + \sum_{L=1}^{2} \sum_{i=1}^{3} \sum_{q=1}^{3} \beta_q(L) \sum_{x,\mu<\nu} \cos q f_{\mu\nu}^i(L) \qquad (36)$$

where $L$ refers to Wilson loop size: $\cos q f_{\mu\nu}(L)$ is an $L \times L$ plaquette in $U(1)$

representation $q$. Table 2 lists the results for ansatz

$$-S_b \equiv \log \delta_\theta + \log \delta_k + \sum_{i=1}^{3}\Big\{-\frac{\kappa}{2}(k^i, k^i) + \sum_{q=1}^{3} \beta_q(1) \sum_{x,\mu<\nu} \cos q f^i_{\mu\nu}(1)\Big\}. \quad (37)$$

$k^i$ in $S_b$ refers to the Toussaint-Degrand $1^3$ monopole current. Examination of Tables 1 and 2 reveals the following:

- $L > 1$ Wilson loops do not contribute significantly to $S_{APQCD}$: $L \geq 2$ Wilson loops have negligibly small couplings in $S_a$ and, further, their presence(absence) in $S_a(S_b)$ does not greatly affect the values of $\beta_q(1)$ in $S_a$ and $S_b$.

- $\beta_{1,2}$ are nonzero, but $\beta_{q\geq 3}$ are too small to be resolved. $q = 1$ operators are dominant.

- $\kappa$ is small; its absence(presence) in $S_a(S_b)$ does not greatly affect the values of $\beta_q(1)$ in $S_a$ and $S_b$.

Thus, except for a small $q = 2$ correction $S_0$ would seem to be a close approximation to $S_{APQCD}$. However, there are two very serious, unresolved problems. Firstly, simulations of $S_b$ with Table 2 couplings indicate that it does *not* reproduce APQCD expectations values: $S_b$ at $\beta_1 = .82$, $\beta_2 = .07$, and $\kappa = 0$ has average plaquette $\langle \cos f^i_{\mu\nu}(1) \rangle = .80(.001)$ and monopole density $\langle |k^i_4| \rangle = .0007(.0002)$—a dramatic discrepancy with APQCD at $\beta_{QCD} = 5.7$, which has average plaquette $.71(.001)$ and monopole density $.048(.001)$. Secondly, APQCD would not be confining if one believes Table 2; simulations indicate that $S_b$ is not confining above $\beta_1 = .585(.05)$ when $\beta_2 = .07$ and $\kappa = 0$. The inability of these microcanonical demon results to reproduce

APQCD tells us that $S_{APQCD}$ has a class of important operators we have neglected. Such operators may involve, for example, nonlocal interactions between pairs of Wilson loops which can arise from integrating out the $A^{ch}$ and $\Lambda$ fields.

At this writing I suspect the problem is the following. It is known that at $\beta_{QCD} = 5.7$ the lattice spacing is $a \sim 1 GeV^{-1}$. As $a$ is shorter than $1/M_{ch} \sim 2 GeV^{-1}$ we cannot properly regard $A^{ch}$ as being "heavy" relative to $S_a$ and $S_b$, which involve $1 \times 1 \equiv a \times a$ plaquettes. $a \times a$ plaquettes would have nonlocal interactions arising from the propagation of virtual $A^{ch}$ loops. A possible remedy is to reformulate $S_{APQCD}$ entirely in terms of $L > 1/(aM_{ch})$ Wilson loops. Its disadvantage is that the relation of such an action to pointlike $1^3$ monopoles—which are known to scale in MA gauge—is complicated; one cannot easily write down a relation for it like (34). This approach is currently under investigation.

## 4. Acknowledgements

I am indebted to Dick Haymaker, Misha Polikarpov, Greg Poulis, Howard Trottier, and Richard Woloshyn for stimulating discussions, and to Mike Creutz for comments about the microcanonical demon. It is a pleasure to thank Professor Faqir Khanna and Ms. Audrey Schaapman of the Lake Louise Winter Institute for the opportunity to present my work at such an enjoyable workshop. Computing was done at the NERSC Supercomputer Center. The author is supported by DOE grant DE-FG05-91ER40617.